\documentclass{elsart}
\usepackage{amsfonts}
\usepackage{amssymb}
\usepackage{amsmath}
\usepackage[dvips]{graphicx,psfrag,overpic,color}

\definecolor{dgreen}{rgb}{0,.6,0}

\begin{document}

\begin{frontmatter}

\title{Security analysis of communication system based on the
synchronization of different order chaotic systems}
\author[Spain]{Gonzalo Alvarez\corauthref{corr}},
\author[Spain]{Luis Hern\'{a}ndez},
\author[Spain]{Jaime Mu\~{n}oz},
\author[Spain]{Fausto Montoya} and
\author[China]{Shujun Li}

\corauth[corr]{Corresponding author: Email: gonzalo@iec.csic.es}

\address[Spain]{Instituto de F\'{\i}sica Aplicada, Consejo Superior de
Investigaciones Cient\'{\i}ficas, Serrano 144, 28006--Madrid,
Spain}
\address[China]{Department of Electronic and Information Engineering, Hong Kong Polytechnic University, Hung Hom, Kowloon, Hong Kong SAR, China}

\begin{abstract}
This letter analyzes the security weakness of a recently proposed
communication method based on chaotic modulation and masking using
synchronization of two chaotic systems with different orders. It
is shown that its application to secure communication is unsafe,
because it can be broken in two different ways, by high-pass
filtering and by reduced order system synchronization, without
knowing neither the system parameter values nor the system key.


\end{abstract}

\end{frontmatter}

\bibliographystyle{elsart-num}

\sloppy

\section{Introduction}\label{intro}

In recent years, a considerable effort has been devoted to extend
the chaotic communication applications to the field of secure
communications. It has been noticed that there exists an
interesting relationship between chaos and cryptography: many
properties of chaotic systems have their corresponding
counterparts in traditional cryptosystems, such as:

\begin{itemize}
\item Ergodicity and Confusion: The output has the same distribution for any
input.

\item Sensitivity to initial conditions/control parameter and Diffusion with a small change in
the plaintext/secret key: A small deviation in the input can cause
a large change at the output.

\item Mixing property and Diffusion with a small change in one plain-block of the whole
plaintext: A small deviation in the local area can cause a large
change in the whole space.

\item
Deterministic dynamics and Deterministic pseudo-randomness: A
deterministic process can cause a random-like (pseudo-random)
behavior.

\item
Structure complexity and Algorithm (attack) complexity: A simple
process has a very high complexity.
\end{itemize}

As a result of investigating the above relationships, a rich
variety of chaos-based cryptosystems for end-to-end communications
have been proposed
\cite{Cuomo93a,Cuomo93b,Wu93,Lozi93,Li01d,Yang04}, some of them
fundamentally flawed by a lack of robustness and security
\cite{short94,zhou97a,alvarez03,alvarez03b,Li03,Li03a,alvarez04}.

Most analog chaos-based cryptosystems are secure communication
schemes designed for noisy channels, based on the technique of
chaos synchronization, first shown by Pecora and Carrol
\cite{pecora90}.

Reduced order synchronization is a new interesting topic which has
recently drawn attention from several researchers
\cite{Xiaofeng00,Femat02,Wang01}. In \cite{Femat02} it is shown
that second order driven oscillators can be synchronized with
canonical projection of a higher order chaotic system by means of
non-linear feedback.

In a recent paper Bowong proposed a scheme based on reduced order
synchronization and feedback with application to secure
communications \cite{Bowong04}. The transmitter is a four-order
chaotic oscillator, modulated by the plaintext and whose output is
added to the plaintext as a masking signal. The receiver consists
of a Duffing second-order system, that is enslaved to the
transmitter by means of non-linear feedback of the error.

Bowong presented two examples based on the plaintext modulation of
a chaotic oscillator and subsequent additive masking of the
plaintext with the oscillator signal. The following equation
system defines the transmitter operation:
\begin{align}
  \dot {x}_{1m} =&~ x_{2m},\nonumber\\
  \dot {x}_{2m} =&~ -cx_{2m}-d^2(x_{1m}-x_{3m}-x_{1m}x_{3m})+ u(t),\label{master}\\
  \dot {x}_{3m} =&~ x_{4m},\nonumber\\
  \dot {x}_{4m} =&~ -ax_{4m}-x_{3m}-b^2x^3_{3m}-f(x_{1m}+x^2_{1m}/2)+e(t),\nonumber\\
            y_T =&~ x_{1m}+u(t)\nonumber,
\end{align}
with parameter values:
\begin{equation}
a=0.03, b=1, c=0.3, d=0.985, f=0.1, \label{param1}
\end{equation}
where the function $e(t)$ was specified as $e(t)=0.7\cos(t)$ and
being the term $u(t)$ the plaintext message. The signal
$y_T=x_{1m}+u(t)$ constitutes the transmitted ciphertext to the
receiver end. Actually, the above Eq.~(\ref{master}) is written as
it should be, in spite of the erroneous formulation given in (14)
of \cite[\S4]{Bowong04}.

The receiver was constructed as follows:
\begin{align*}
  \dot {x}_{1s}  =&~x_{2s},\\
  \dot {x}_{2s}  =&~ -\lambda x_{2s}-\omega_0^2 x_{1s}-\gamma x^3_{1s}
                  +K_1 \cos(\omega_1 t+\theta_1)
                  +K_2 \cos(\omega_2 t+\theta_2) +\upsilon ,\\
  y_s =&~ x_{1s},
\end{align*}
with parameter values:
\begin{equation}
\lambda=1, \omega_0=10, \gamma=100, K_1=K_2=1, \omega_1=2,
\omega_2=4, \theta_1=\theta_2=0, \label{param2}
\end{equation}
and where $\upsilon$ is the feedback control law which
forces the error $e=x_{1s}-x_{1m}$ to converge exponentially to zero as
$t\rightarrow\infty$.

The retrieved plaintext $\hat{u}(t)$ is calculated as the
difference between the ciphertext and the output of the reduced
order system $\hat{u}(t)=y_T - y_s=u(t)-e$. It was shown that the
transmitter-receiver system was capable of accurately retrieving
the plaintext after an initial synchronization period of 10
seconds. Afterwards it was claimed that the system can be used for
secure communications and some examples are provided.

In this letter it is shown that the proposed cryptosystem is
insecure and two different procedures to break it are also
presented: by high-pass filtering and by means of a simple reduced
order intruder receiver.

\section{Missing security analysis and system key specification}

In \cite{Bowong04}, the author asserted that the scheme is
applicable to secure communication. However, no analysis of
security was included to support this claim. Furthermore, there is
no mention of the secret key, when it is well known that a secure
communication system cannot exist without a key. In
\cite{Bowong04} it is not considered whether there should be a key
in the proposed system, what it should consist of, what the
available key space would be (how many different keys exist in the
system), what precision to use, and how it would be managed. None
of these elements should be neglected when describing a secure
communication system \cite{AlvarezLi03}.

Moreover, being the transmitter and the receiver implemented with
different kind of systems, it is not explained how the encryption
keys, if any, may be related to the corresponding decryption keys.
Usually, in many chaotic cryptosystems the system parameters play
the role of key, but it is not the case in \cite{Bowong04},
because the transmitter and the receiver do not make use of the
same parameters.

\section{Plaintext retrieval by a filtering attack}

It was supposed for some time that chaotic masking was an adequate
means for secure transmission, because chaotic systems present
some properties as sensitive dependence on parameters and initial
conditions, ergodicity, mixing, and dense periodic points. These
properties make them similar to pseudorandom noise
\cite{Devaney92}, which has been used traditionally as a masking
signal for cryptographic purposes. The basic fundamental
requirement of the pseudorandom noise used in cryptography is that
its spectrum should be infinitely broad, flat and of much higher
power density than the signal to be concealed. In other words, the
plaintext power spectrum should be effectively buried into the
pseudorandom noise power spectrum.

The secure application proposed in \cite{Bowong04} does not
satisfy this condition. On the contrary, the spectrum of the
signal generated by the chaotic oscillator is of narrow band,
decaying very fast with increasing frequency, showing a power
density much lower than the plaintext at the plaintext frequencies
used. Hence it can not cope with a filtering attack intended to
separate the masking signal and the plaintext.

\begin{figure}[h]
\begin{center}
\includegraphics[scale=0.5]{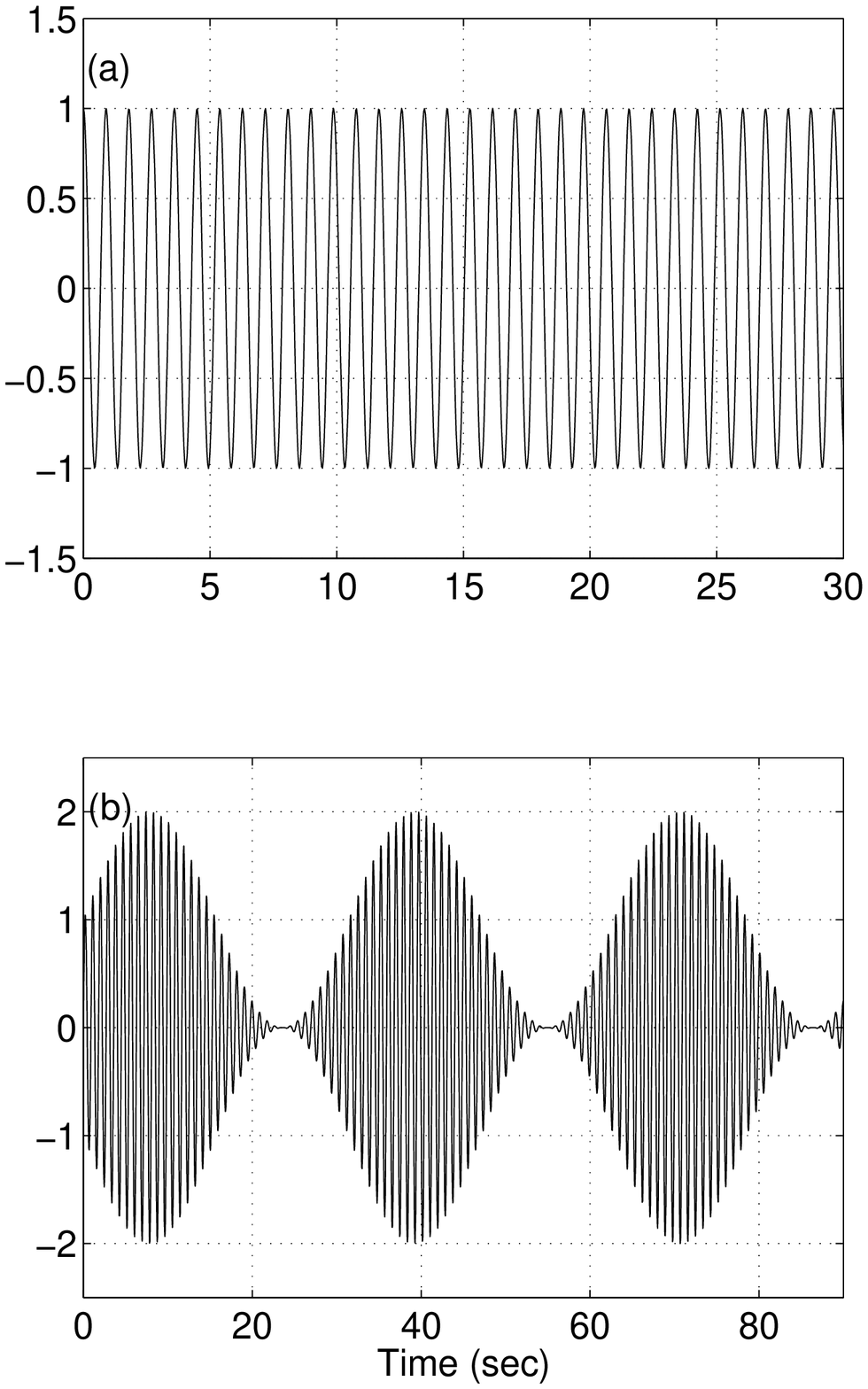}\\
\caption{Plaintext examples described in \cite{Bowong04}: (a)
$u_1(t) = \cos (7 t)$; (b) $u_2(t) = (1+\sin (0.2 t))\cos (7 t)$.}
\label{fig:plaintext}
\end{center}
\end{figure}

To illustrate this fact we consider the the two examples in
\cite[\S4]{Bowong04} corresponding to the following plaintexts:
\begin{align*}
  u_1(t) &=\cos (7 t),\\
  u_2(t) &=(1+\sin (0.2 t)) \cos (7 t),
\end{align*}
whose waveforms are illustrated in Fig.~\ref{fig:plaintext}.

\begin{figure}[h]
\begin{center}
\includegraphics[scale=0.5]{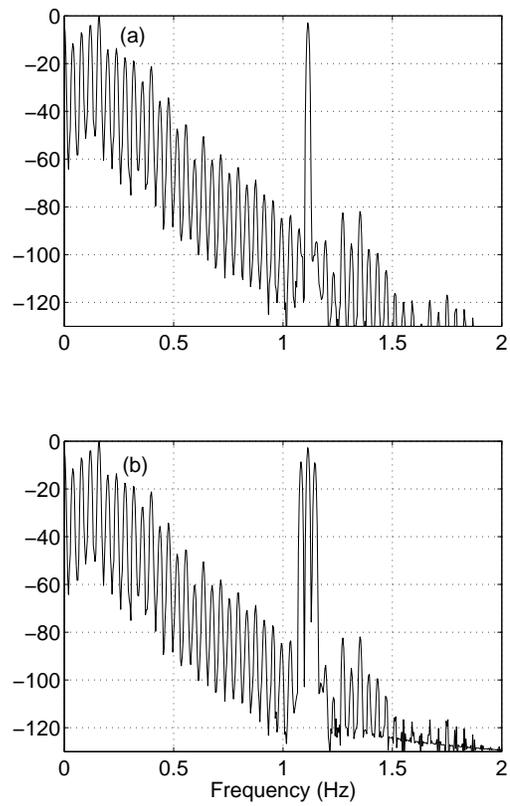}
\caption{Logarithmic power spectra of the ciphertexts $y_{T1}$ and
$y_{T2}$: (a) corresponding to the plaintext $u_1(t)=\cos (7 t)$;
(b) corresponding to the plaintext $u_2(t)=(1+\sin (0.2 t)) \cos
(7 t).$} \label{fig:spectrum1}
\end{center}
\end{figure}

The transmitter proposed  in~\cite{Bowong04} was simulated with a
four-order Runge-Kutta integration algorithm in MATLAB 6.5, with a
step size of $10^{-3}$. Fig.~\ref{fig:spectrum1} illustrates the
logarithmic power spectra, as a function of frequency, of the
ciphertexts $y_{T1}$ and $y_{T2}$ when the plaintext signals
$u_1(t)$ and $u_2(t)$ are encrypted, respectively, with the same
parameter values previously described in~(\ref{param1}). The power
spectra were calculated using a 8192-point Discrete Fourier
Transform with a sampling frequency of 32~Hz; previously, the
analyzed signal segments were multiplied by a 4-term
Blackman-Harris window \cite{Harris78}, to avoid aliasing
artifacts.

It can be seen in both examples that the plaintext signal
components clearly emerge at 1.114 Hz over the background noise
created by the chaotic oscillator, with a power of
$-3~\textrm{db}$, relative to the maximum power of the ciphertext
spectrum, while the power density of the ciphertext, at
neighboring frequencies, falls below $-80~ \textrm{db}$.

\begin{figure}[h]
\begin{center}
\includegraphics[scale=0.5]{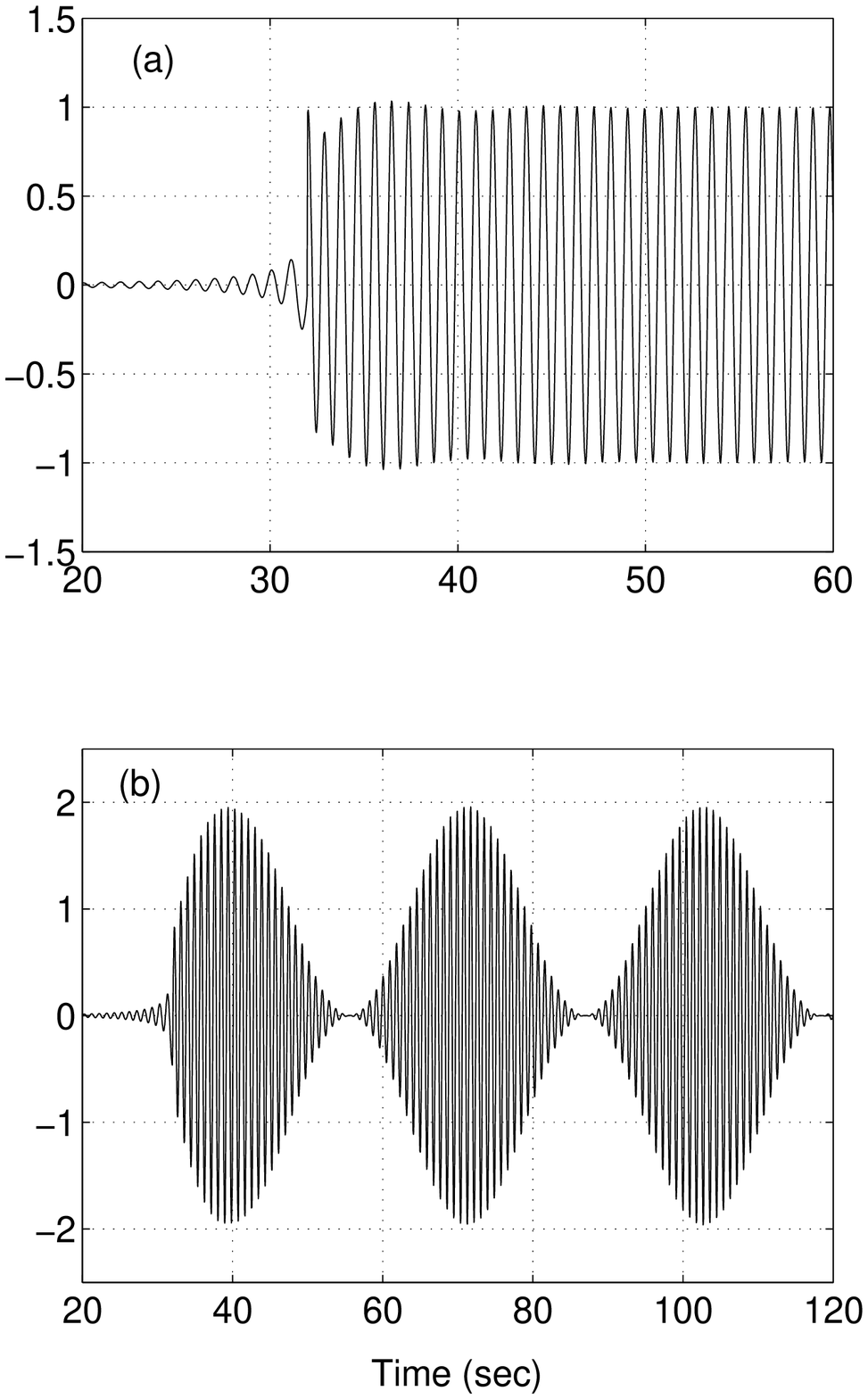}
\caption{Retrieved texts $\hat{u}_1(t)$ and $\hat{u}_2(t)$, by
high pass filtering of the two plaintext examples: (a)
$u_1(t)=\cos (7 t)$; (b) $u_2(t)=(1+\sin (0.2~t))~\cos (7~t)$.}
\label{fig:retrieved1}
\end{center}
\end{figure}

The chaotic receiver of \cite{Bowong04} was not used to recover
the plaintext. Instead, the ciphertext was high-pass filtered to
eliminate the chaotic masking component while retaining the
plaintext information. The result is illustrated in
Fig.~\ref{fig:retrieved1}. Comparing the result with the plaintext
displayed in Fig.~\ref{fig:plaintext}, it can be appreciated the
good estimation of the plaintexts after an initial delay of
approximately 29 seconds, due to the filter delay. The filter
employed was a 2048 samples finite impulse response digital one,
with a cut-off frequency of 1~Hz.

Note that this is the hardest case an attacker can face from the
point of view of plaintext frequency, because for higher sound
frequencies the spectrum of the background noise created by the
chaotic oscillator is even lower.

This plaintext recovering method works equally well for different
parameters values of the transmitter, because the maximum power
components of its spectrum are concentrated in the frequency range
between 0 and 0.3~Hz for all parameter values.

\begin{figure}[h]
\begin{center}
\includegraphics[scale=0.5]{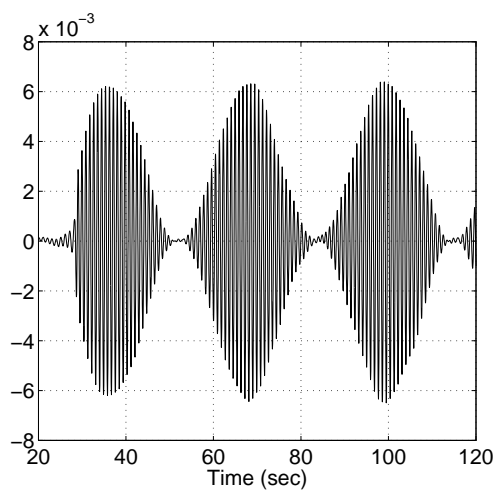}
\caption{Retrieved text $\hat{u}_3(t)$ by high-pass filtering of
the ciphertext $y_T=x_{1m}+u_3(t)$, corresponding to the low power
level plaintext $u_3(t)=0.0032(1+\sin (0.2 t)) \cos (7 t)$.}
\label{fig:retrieved3}
\end{center}
\end{figure}

Thanks to the big separation between the plaintext frequency and
the high amplitude components of the masking chaotic signal, our
method works equally well with plaintext signals of much lower
amplitudes than the plaintexts $u_1(t)$ and $u_2(t)$ of the
examples described in \cite{Bowong04}. For instance, we present in
Fig.~\ref{fig:retrieved3} the retrieved text corresponding to the
plaintext $u_3(t)=0.0032 (1+\sin (0.2 t)) \cos (7 t)$, that has a
power level of -50db with respect to $u_2(t)$; but, as can be
seen, the retrieved signal waveform is still perfectly preserved.

\section{Plaintext retrieval by reduced order system synchronization}
As mentioned in Sec.~\ref{intro}, many secure communication
systems based on chaotic modulation and masking have been proposed
in the past. In any of them the knowledge of the transmitter
parameter values was mandatory to operate the receiver, since they
played the role of system key. Some of them were compromised
because it was possible to recover the system parameters carrying
out an elaborate ciphertext signal analysis.

But the chaotic transmitter and the modulation and masking
procedure described in \cite{Bowong04} was designed in such a way
as to enable the plaintext retrieval with a reduced order receiver
that did not even required the knowledge of any transmitter
parameters. As a true encryption mechanism must necessarily make
use of a key, this communication system may be only envisaged as
an ordinary codification system, rather than a secure one, because
the only required knowledge to recover the plaintext message is
the receiver structure.

Moreover, it is possible to implement a whole family of
alternative receivers to the one proposed in \cite{Bowong04}.
Hence, a determined eavesdropper, still ignoring the precise
structure of the transmitter nor its design parameters, may
implement an alternative intruder receiver of its own design, also
based on reduced order synchronization and feedback, capable of
retrieving the ciphertext just as well as the authorized one.

To demonstrate this threat we have developed an extremely simple
intruder receiver of order two, with linear feedback, constructed
as follows:
\begin{align*}
  \dot {x}_{1s}  =& ~x_{2s},\\
  \dot {x}_{2s}  =& ~100~ \hat{u}(t)- x_{2s},
\end{align*}
being  $\hat{u}(t)=y_T - {x}_{1s}$ the retrieved plaintext. The
initial conditions were arbitrarily chosen as
\[
x_{1m}(0)=x_{2m}(0)=x_{3m}(0)=x_{4m}(0)=0, x_{1s}(0)=0.1,
x_{2s}(0)=1.
\]

\begin{figure}[h]
\begin{center}
\includegraphics[scale=0.5]{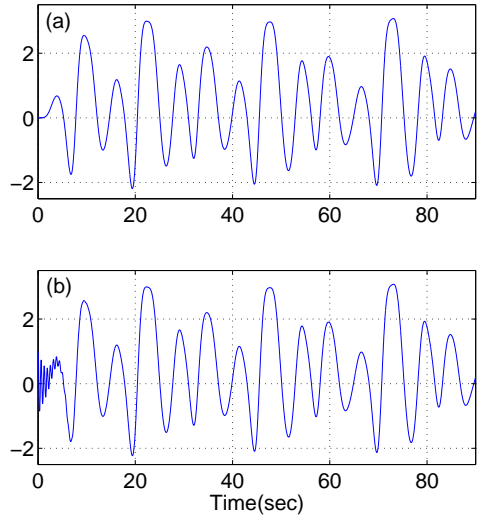}
\caption{Synchronism between transmitter and intruder receiver,
when no plaintext signal is present: (a) transmitter variable
${x}_{1m}$ v. time; (b) receiver variable ${x}_{1s}$ v. time.}
\label{fig:synchro}
\end{center}
\end{figure}

\begin{figure}[h]
\begin{center}
\includegraphics[scale=0.5]{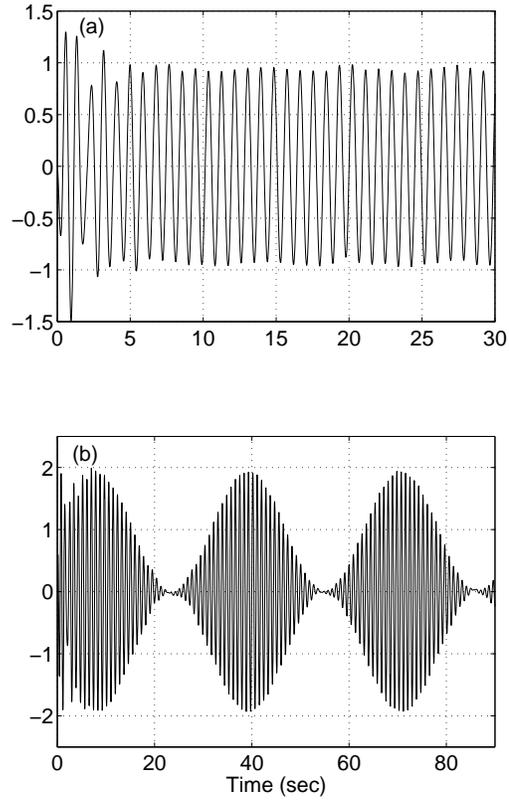}
\caption{Retrieved texts $\hat{u}_1(t)$ and $\hat{u}_2(t)$, by our
intruder receiver for the two plaintext examples: (a) $u_1(t)=\cos
(7 t)$; (b) $u_2(t)=(1+\sin (0.2~t))~\cos (7~t)$.}
\label{fig:retrieved4}
\end{center}
\end{figure}

This simple receiver may decrypt the ciphertext as well as the
receiver proposed in \cite{Bowong04}. Fig.~\ref{fig:synchro}
illustrates the perfect synchronism between the transmitter
variable ${x}_{1m}$ and our intruder receiver variable ${x}_{1s}$,
attained after a transient of 4 seconds, when no plaintext signal
is present. The efficiency as intruder decoder is illustrated in
Fig.~\ref{fig:retrieved4}, where the retrieved $\hat{u}_1(t)$ and
$\hat{u}_2(t)$ texts corresponding to the plaintexts $u_1(t)=\cos
(7 t)$ and $u_2(t)=(1+\sin (0.2~t))~\cos (7~t)$ are shown,
comparing with the plaintexts illustrated in
Fig.~\ref{fig:plaintext} it can be appreciated the perfect
decoding after a short initial transient.

\section{Conclusion}
In summary, the chaotic masking cryptosystem proposed in
\cite{Bowong04} is rather weak, since it can be broken in two
different ways, without knowing the system parameters nor it
detailed structure: by high pass filtering and by an intruder
receiver based on reduced order synchronization and feedback.
There is no mention about what the key is, nor what the key space
is, a fundamental aspect in every secure communication system. The
total lack of security discourages the application of this
synchronization scheme to secure applications.

\ack{This work was supported by Ministerio de Ciencia y
Tecnolog\'{\i}a of Spain, research grant SEG2004-02418. We thank
the anonymous reviewer for his valuable suggestions.}

\bibliography{criptsec,caos90tic04,fmontoya}

\newpage

\section*{Figures}

\end{document}